%#!tex v2-thooft.tex
%% Last Modified: Sun Aug 13 12:02:39 2006.

%%%%%%%%%%%%%%%%%%%%%%%%%%%%%%%%%%%%%%%%%%%%%%%%%%%%%%%%%%%%%%%%%
%                                                               %
%                          NS5                                  %
%                                                               %
%                  Kazumi Okuyama (UBC)                         %
%                                                               %
%%%%%%%%%%%%%%%%%%%%%%%%%%%%%%%%%%%%%%%%%%%%%%%%%%%%%%%%%%%%%%%%%

%\input harvmac

%\def\mydraft{label}

\input lanlmac
\input amssym
\input epsf

%Macro for figure
\newcount\figno
\figno=0
\def\fig#1#2#3{
\par\begingroup\parindent=0pt\leftskip=1cm\rightskip=1cm\parindent=0pt
\baselineskip=13pt
\global\advance\figno by 1
\midinsert
\epsfxsize=#3
\centerline{\epsfbox{#2}}
\vskip 12pt
%\centerline{{\bf Fig. \the\figno:~~} #1}\par
{\bf Fig. \the\figno:~~} #1 \par
\endinsert\endgroup\par
}
\def\figlabel#1{\xdef#1{\the\figno}}
%%                              TABLEAUX.TEX
%%      This  macro file is for producing a ``Young Tableau'' which is
%%      an array of little squares sometimes used in mathematical physics.
%%      For instance, the command $\tableau{6 3 2}$ will produce a tableau
%%      with 6 squares in the top row, 3 in the next, and 2 in the last.
%%                                  OOOOOO
%%      This tableau will look like OOO    but made of squares instead of
%%           O's.
%%                                  OO
%%      Any number of rows may be present, each having a nonzero number of
%%      squares.
%%
%%      A tableau is math mode material, so use $ or $$ to enclose it.
%%
%%      The size and line-thickness of the little boxes are controlled by
%%  the
%%      dimension parameters --
%%              \tableauside=1.0ex              %(size)
%%              \tableaurule=0.4pt              %(line-thickness)
%%      Change them if you want.
%%
%%                                                      -- Doug Eardley
%%   9/19/8
%%
%%
\newdimen\tableauside\tableauside=1.0ex
\newdimen\tableaurule\tableaurule=0.4pt
\newdimen\tableaustep
\def\phantomhrule#1{\hbox{\vbox to0pt{\hrule height\tableaurule
width#1\vss}}}
\def\phantomvrule#1{\vbox{\hbox to0pt{\vrule width\tableaurule
height#1\hss}}}
\def\sqr{\vbox{%
  \phantomhrule\tableaustep

\hbox{\phantomvrule\tableaustep\kern\tableaustep\phantomvrule\tableaustep}%
  \hbox{\vbox{\phantomhrule\tableauside}\kern-\tableaurule}}}
\def\squares#1{\hbox{\count0=#1\noindent\loop\sqr
  \advance\count0 by-1 \ifnum\count0>0\repeat}}
\def\tableau#1{\vcenter{\offinterlineskip
  \tableaustep=\tableauside\advance\tableaustep by-\tableaurule
  \kern\normallineskip\hbox
    {\kern\normallineskip\vbox
      {\gettableau#1 0 }%
     \kern\normallineskip\kern\tableaurule}% 
  \kern\normallineskip\kern\tableaurule}}
\def\gettableau#1 {\ifnum#1=0\let\next=\null\else
  \squares{#1}\let\next=\gettableau\fi\next}

\tableauside=1.0ex
\tableaurule=0.4pt

%Macro

\def\th{\theta}

\def\Tr{{\rm Tr}}
\def\hf{{1\over 2}}
\def\qu{{1\over 4}}

\def\o{\over}

\def\til#1{\widetilde{#1}}

\def\del{\partial}

\def\bra{\langle}
\def\ket{\rangle}
\def\lf{\left}
\def\ri{\right}
\def\riya{\rightarrow}

\def\lrya{\leftrightarrow}

\def\la{\lambda}

\def\h#1{\widehat{#1}}

\def\ga{\gamma}

\def\al{\alpha}

\def\rt#1{\sqrt{#1}}

\def\sitarel#1#2{\mathrel{\mathop{\kern0pt #1}\limits_{#2}}}
\def\uerel#1#2{{\buildrel #1 \over #2}}

\def\Li{{\rm Li}}

%\newsec{References}
\lref\DrukkerRR{
  N.~Drukker and D.~J.~Gross,
  ``An Exact Prediction Of N = 4 Susym Theory For String Theory,''
  J.\ Math.\ Phys.\  {\bf 42}, 2896 (2001)
  [arXiv:hep-th/0010274].
  %%CITATION = HEP-TH 0010274;%%
}
\lref\EricksonAF{
  J.~K.~Erickson, G.~W.~Semenoff and K.~Zarembo,
  ``Wilson loops in N = 4 supersymmetric Yang-Mills theory,''
  Nucl.\ Phys.\ B {\bf 582}, 155 (2000)
  [arXiv:hep-th/0003055].
  %%CITATION = HEP-TH 0003055;%%
}
\lref\Buchholz{
H. Buchholz, {\it The Confluent Hypergeometric Function}
(1969), Springer-Verlag.
}
\lref\Abad{
J. Abad and J. Sesma,
``Computation of the Regular Confluent Hypergeometric Function,''
Mathematica Journal {\bf 5}, 4 (1995).
}
\lref\tHooftJZ{
  G.~'t Hooft,
  ``A Planar Diagram Theory For Strong Interactions,''
  Nucl.\ Phys.\ B {\bf 72}, 461 (1974).
  %%CITATION = NUPHA,B72,461;%%
}
\lref\BianchiGZ{
  M.~Bianchi, M.~B.~Green and S.~Kovacs,
   ``Instanton corrections to circular Wilson loops in N = 4 supersymmetric
  Yang-Mills,''
  JHEP {\bf 0204}, 040 (2002)
  [arXiv:hep-th/0202003].
  %%CITATION = HEP-TH 0202003;%%
}
\lref\GopakumarKI{
  R.~Gopakumar and C.~Vafa,
  ``On the gauge theory/geometry correspondence,''
  Adv.\ Theor.\ Math.\ Phys.\  {\bf 3}, 1415 (1999)
  [arXiv:hep-th/9811131].
  %%CITATION = HEP-TH 9811131;%%
}
\lref\BerensteinIJ{
  D.~Berenstein, R.~Corrado, W.~Fischler and J.~M.~Maldacena,
   ``The operator product expansion for Wilson loops and surfaces in the  large
  N limit,''
  Phys.\ Rev.\ D {\bf 59}, 105023 (1999)
  [arXiv:hep-th/9809188].
  %%CITATION = HEP-TH 9809188;%%
}
\lref\DrukkerZQ{
  N.~Drukker, D.~J.~Gross and H.~Ooguri,
  ``Wilson loops and minimal surfaces,''
  Phys.\ Rev.\ D {\bf 60}, 125006 (1999)
  [arXiv:hep-th/9904191].
  %%CITATION = HEP-TH 9904191;%%
}
\lref\MaldacenaIM{
  J.~M.~Maldacena,
  ``Wilson loops in large N field theories,''
  Phys.\ Rev.\ Lett.\  {\bf 80}, 4859 (1998)
  [arXiv:hep-th/9803002].
  %%CITATION = HEP-TH 9803002;%%
}
\lref\ReyIK{
  S.~J.~Rey and J.~T.~Yee,
   ``Macroscopic strings as heavy quarks in large N gauge theory and  anti-de
  Sitter supergravity,''
  Eur.\ Phys.\ J.\ C {\bf 22}, 379 (2001)
  [arXiv:hep-th/9803001].
  %%CITATION = HEP-TH 9803001;%%
}
\lref\WittenHF{
  E.~Witten,
  ``Quantum field theory and the Jones polynomial,''
  Commun.\ Math.\ Phys.\  {\bf 121}, 351 (1989).
  %%CITATION = CMPHA,121,351;%%
}
\lref\BershadskyCX{
  M.~Bershadsky, S.~Cecotti, H.~Ooguri and C.~Vafa,
   ``Kodaira-Spencer theory of gravity and exact results for quantum string
  amplitudes,''
  Commun.\ Math.\ Phys.\  {\bf 165}, 311 (1994)
  [arXiv:hep-th/9309140].
  %%CITATION = HEP-TH 9309140;%%
}
\lref\OoguriGX{
  H.~Ooguri and C.~Vafa,
  ``Worldsheet Derivation Of A Large N Duality,''
  Nucl.\ Phys.\ B {\bf 641}, 3 (2002)
  [arXiv:hep-th/0205297].
  %%CITATION = HEP-TH 0205297;%%
}
\lref\MaldacenaRE{
  J.~M.~Maldacena,
  ``The large N limit of superconformal field theories and supergravity,''
  Adv.\ Theor.\ Math.\ Phys.\  {\bf 2}, 231 (1998)
  [Int.\ J.\ Theor.\ Phys.\  {\bf 38}, 1113 (1999)]
  [arXiv:hep-th/9711200].
  %%CITATION = HEP-TH 9711200;%%
}
\lref\AharonyTI{
  O.~Aharony, S.~S.~Gubser, J.~M.~Maldacena, H.~Ooguri and Y.~Oz,
  ``Large N field theories, string theory and gravity,''
  Phys.\ Rept.\  {\bf 323}, 183 (2000)
  [arXiv:hep-th/9905111].
  %%CITATION = HEP-TH 9905111;%%
}
\lref\MarinoFK{
  M.~Marino,
   ``Chern-Simons Theory, Matrix Integrals, And Perturbative Three-Manifold
  Invariants,''
  Commun.\ Math.\ Phys.\  {\bf 253}, 25 (2004)
  [arXiv:hep-th/0207096].
  %%CITATION = HEP-TH 0207096;%%
}
\lref\AganagicWV{
  M.~Aganagic, A.~Klemm, M.~Marino and C.~Vafa,
  ``Matrix model as a mirror of Chern-Simons theory,''
  JHEP {\bf 0402}, 010 (2004)
  [arXiv:hep-th/0211098].
  %%CITATION = HEP-TH 0211098;%%
}
\lref\DrukkerKX{
  N.~Drukker and B.~Fiol,
  ``All-genus calculation of Wilson loops using D-branes,''
  JHEP {\bf 0502}, 010 (2005)
  [arXiv:hep-th/0501109].
  %%CITATION = HEP-TH 0501109;%%
}
%%%%%%%%%%%%%%%%%%%%%%%%%%%%%%%%%%%%%%%%%%%%%%%%%%%%%%%%%%%%%%%%%
%                      Title Page                               %
%%%%%%%%%%%%%%%%%%%%%%%%%%%%%%%%%%%%%%%%%%%%%%%%%%%%%%%%%%%%%%%%%
\Title{             
                                             \vbox{
                                             \hbox{hep-th/0607131}}}
{\vbox{
\centerline{'t Hooft Expansion of 1/2 BPS Wilson Loop}
}}

\vskip .2in

\centerline{Kazumi Okuyama}

\vskip .2in

%\vskip 2cm
\centerline{Department of Physics and Astronomy, 
University of British Columbia} 
\centerline{Vancouver, BC, V6T 1Z1, Canada}
\centerline{\tt kazumi@phas.ubc.ca}
\vskip 3cm
\noindent

%%abstract
We revisit the 't Hooft expansion of 1/2 BPS circular Wilson loop
in ${\cal N}=4$ SYM studied by Drukker and Gross in hep-th/0010274.
We find an interesting recursion relation which relates different
number of holes on the worldsheet.
We also argue that we can turn on the string coupling 
by applying a certain integral transformation to the planar result.

\Date{July 2006}

\vfill
\vfill

\newsec{Introduction}
Even after the revolutions in string theory,
we still lack enough control over the string coupling dependence of various
amplitudes.
Notable exceptions are the $c\leq1$ non-critical
strings and the topological strings.
For the latter case, all genus results are available for some models.
Moreover, there is a deep underlying structure
in the genus expansion which enables us to determine 
the amplitudes recursively \BershadskyCX.
It is extremely interesting to find a similar relation
in the non-topological setup, but in general it is a formidable task.

However, there exists an all order result in our familiar
AdS/CFT context: the expectation value of 1/2 BPS circular
Wilson loop in $U(N)$ ${\cal N}=4$ SYM \DrukkerRR.
In the string theory side, Wilson loop is described by a string
worldsheet ending on the loop at the boundary of
AdS, and its expectation value is given by the worldsheet action for
the minimal surface
\refs{\ReyIK,\MaldacenaIM,\DrukkerZQ}.
On the Yang-Mills side, 1/2 BPS circular Wilson loop has the form
\eqn\TrPW{
W=\lf\bra{1\o N}\Tr P\exp\lf(\oint dt(iA_\mu \dot{x}^\mu+
\th^i\Phi_i|\dot{x}|)\ri)\ri\ket
}
where $\th^i$ is a constant unit vector in ${\Bbb R}^6$ and $x^\mu(t)$
is a circular loop in ${\Bbb R}^4$.
In a beautiful paper \EricksonAF, it is realized that
summing over
the rainbow diagrams boil down to the
following $N\times N$ Hermitian matrix model
\eqn\matint{
W={1\o Z}\int dM
\exp\lf(-{2\o g_{\rm YM}^2}\Tr M^2\ri)
{1\o N}\Tr\,e^M~.
}
In \DrukkerRR\ it is further argued that this matrix model actually
gives 
all order result of \TrPW\ in perturbation theory, 
up to possible instanton corrections \BianchiGZ.

In this paper, we will study the 't Hooft expansion \tHooftJZ\ of
circular Wilson loop
\eqn\thooftexp{
W(\la,g_s)=\sum_{h,\ell=0}^\infty {\cal W}_{h,\ell}\la^h g_s^{2\ell}
}
in terms of the string coupling  $g_s\sim g_{\rm YM}^2$ and
the 't Hooft coupling $\la=g_{\rm YM}^2N$. As reviewed
nicely in  \OoguriGX, a Feynman diagram of Yang-Mills theory
is reorganized as a Riemann surface
of $h$ holes and $\ell$ handles.
Note that the $g_s$-expansion of Wilson loop starts with the order
$g_s^0$,
since the trace in \TrPW\ is normalized by $1/N$.
At least perturbatively, the double summation in \thooftexp\
can be performed
by either summing over $h$ first or
summing over $\ell$ first.
The first choice leads to the usual genus expansion of
$W$. In section 2, we will consider how to
find the $\ell$-loop term systematically.
The second choice leads to the expansion
of $W$ in terms of the number of holes.
In section 3, we will find a recursion relation between
the $h$-hole amplitude and the $(h+1)$-hole amplitude.
In both cases, we find a curious property of circular Wilson loop:
we can turn on the coupling from zero-coupling
by a certain operation:
\eqn\Wturnon{\eqalign{
W(0,g_s)&\longrightarrow W(\la,g_s)~, \cr
W(\la,0)\,& \longrightarrow W(\la,g_s)~.
}}
In section 4, we will study the relation between $W$ at
$g_s=0$ and $g_s\not=0$, and argue that
they are related by an integral transformation.
Section 5 is the discussions on our findings, 
and some technical details are
collected in the appendices.

\newsec{Genus Expansion of Circular Wilson Loop}
In this section, we will consider the string loop expansion
of circular Wilson loop
\eqn\Wloop{
W(\la,g_s)=\sum_{\ell=0}^\infty W_{\ell}(\la)g_s^{2\ell}~.
}
This problem has been already studied in \DrukkerRR.
However, we find it useful to revisit this problem
since there is a systematics behind this which is not
mentioned in \DrukkerRR.

\subsec{Relation between parameters in YM and String Theory}
Before starting the analysis, let us first review the relation between
the parameters in ${\cal N}=4$ SYM and string theory on $AdS_5\times S^5$
\refs{\MaldacenaRE,\AharonyTI}.
The 't Hooft coupling $\la$ in ${\cal N}=4$ SYM corresponds to
the curvature radius in the string theory side
\eqn\curvR{
(R_{AdS_5})^2=(R_{S^5})^2=\al'\rt{\la}~.
}
In other words, $1/\rt{\la}$ governs the worldsheet sigma-model
corrections. The Yang-Mills coupling $g_{\rm YM}^2$
is related to the string
coupling $g_s$ via the identification of
complex coupling in ${\cal N}=4$ SYM and the axion-dilaton of Type IIB string
\eqn\tauYM{
\tau_{\rm YM}={\th\o2\pi}+i{4\pi\o g_{\rm YM}^2}\quad \Leftrightarrow
\quad\tau_{\rm IIB}=\chi+ie^{-\phi}~.
}
In this paper, we define the string coupling $g_s$ as
\eqn\gsvsgym{
g_s={g_{\rm YM}^2\o4}~.
}
This differs from the usual normalization by a factor of $\pi$.
This factor can be absorbed by a constant shift of the dilaton zero-mode
$\phi_0$. Our normalization is motivated by the fact 
that under the relation \gsvsgym\   the Gaussian
measure of matrix model \matint\ becomes
\eqn\matmeasure{
\int dM\exp\lf(-{2\o g_{\rm YM}^2}\Tr M^2\ri)
=\int dM\exp\lf(-{1\o 2g_s}\Tr M^2\ri)~,
}
which implies that $g_s$ with this normalization
is the canonical loop counting parameter of
the matrix model \matmeasure. In our normalization, 't Hooft coupling
$\la=g_{\rm YM}^2N$ is written as
\eqn\laings{
\la=4g_sN~.
}

\subsec{Expansion in Terms of Buchholz Polynomials}
In \DrukkerRR, the matrix integral
\matint\ was evaluated exactly at finite $N$ and the result
is given  by a Laguerre polynomial
\eqn\Laggs{
W={1\o N}e^{g_s\o2}L^1_{N-1}(-g_s)~.
}
This is also written as a confluent hypergeometric function
\eqn\confW{
W=e^{g_s\o2}{}_1\!F_1(1-N,2\,;-g_s)=e^{- {g_s\o2}}{}_1\!F_1(1+N,2\,;g_s)
~.
}

We would like to recast this into a form of \thooftexp\
in the large $N$ limit with fixed $\la=4g_sN$.
The large $a$ behavior of confluent hypergeometric function
${}_1\!F_1(a,b\,;z)$ with general $b$ has been studied
by mathematicians 
\refs{\Buchholz,\Abad}, so we can borrow their result and 
apply it to our case $b=2$ \confW. The starting point of the analysis 
is the following
contour integral representation of \Laggs
\eqn\Wcont{
W(\la,g_s)=2\oint_{z=0}{dz\o2\pi i}\exp\lf[{\la\o2}z+{g_s\o2}\coth(g_sz)\ri]
~.
}
The equivalence of \Laggs\ and \Wcont\ is shown in appendix A.
Then we add and subtract the single pole of $\coth(g_sz)$ at $z=0$
\eqn\splitH{
W(\la,g_s)=2\oint_{z=0}{dz\o2\pi i}\exp\lf[\hf\lf(\la z+{1\o z}\ri)
+g_sH(g_sz)\ri]~,
}
where $H(x)$ is defined by
\eqn\Hxcoth{
H(x)=\hf\lf(\coth x-{1\o x}\ri)~.
}
Since $H(x)$ is regular at $x=0$, the second factor in
\splitH\ has the Taylor expansion around $z=0$
\eqn\sergsH{
g_sH(g_sz)
={g_s^2\o6}z-{g_s^4\o90}z^3+{g_s^6\o945}z^5-{g_s^8\o9450}z^7+
{g_s^{10}\o93555}z^9+\cdots~.
}
To evaluate the contour integral, we expand the integrand around $z=0$.
Using the generating function of modified Bessel functions
\eqn\modgen{
\exp\lf[{t\o2}\lf(x+{1\o x}\ri)\ri]=\sum_{n=-\infty}^\infty{I_n(t)\o x^n}
~,
}
the first part in \splitH\ has the following Laurent expansion 
\eqn\explaIsum{
\exp\lf[\hf\lf(\la z+{1\o z}\ri)\ri]=
\sum_{k=-\infty}^\infty{\h{I}_k(\la)\o z^k}~,
}
where we introduced a function $\h{I}_k(\la)$ as
\eqn\Ihat{
\h{I}_k(\la)={I_{k}(\rt{\la})\o(\rt{\la})^{k}}~.
}
The second factor of \splitH\ is regular around $z=0$ 
and admits a Taylor expansion.
The coefficient of $x^n$ in the Taylor expansion 
of $e^{aH(x)}$
is known as the Buchholz polynomial $p_n(a)$ \refs{\Buchholz,\Abad}
\eqn\pndef{
\sum_{n=0}^\infty x^np_n(a)~~\uerel{{\rm def}}{=}~~\exp\Big[aH(x)\Big]~.
}
Therefore, the second factor of \splitH\ is expanded as
\eqn\Buchexp{
\exp\Big[g_sH(g_sz)\Big]=\sum_{n=0}^\infty (g_sz)^n p_n(g_s)~.
}
Finally, combining \explaIsum\ and \Buchexp,
the circular Wilson loop \splitH\ is written as 
\eqn\Wbuch{
W(\la,g_s)=2\sum_{n=0}^\infty\h{I}_{n+1}(\la)\,
g_s^np_n(g_s)~.
}

From the definition \pndef, one can easily see that
$p_n(g_s)$ is an $n^{\rm th}$ order polynomial in $g_s$ with fixed parity
$p_n(-g_s)=(-1)^np_n(g_s)$. See appendix B for more
information on $p_n(g_s)$.
To our knowledge, the closed form of $p_n(g_s)$ is not known in
the literature. However, starting from $p_0(g_s)=1$, we can generate 
$p_n(g_s)$
successively using the following recursion relation
\eqn\Pnint{
p_n(g_s)=g_s\int_0^1 dt\,t^{n\o2}\lf[{1\o4}p_{n-1}(g_st)-p_{n-1}''(g_st)\ri]
}
where $p''_{n-1}$ denotes the second derivative of $p_{n-1}$.
For instance, the first few terms are given by
\eqn\pseries{
p_1(g_s)={g_s\o6},\quad p_2(g_s)={g_s^2\o72},\quad
p_3(g_s)={g_s^3\o1296}-{g_s\o90},\quad
p_4(g_s)={g_s^4\o31104}-{g_s^2\o540}~.
}

Although \Wbuch\ is a nice compact expression, we have to rewrite it in
the form of string loop expansion \Wloop.
This is basically a problem of the change of basis
from the polynomials $\{g_s^np_n(g_s)\}_{n=0}^\infty$ to
the monomials $\{g_s^{2\ell}\}_{\ell=0}^\infty$. 
We can easily find a first few terms of $\ell$-loop correction
$W_{\ell}(\la)$ of Wilson loop using \pseries
\eqn\Wels{\eqalign{
W_0(\la)&=2\h{I}_1(\la) \cr
W_1(\la)&={\h{I}_2(\la)\o3} \cr
W_2(\la)&=-{\h{I}_4(\la)\o45}+
{\h{I}_3(\la)\o36} \cr
W_3(\la)&={2\h{I}_6(\la)\o945}-{\h{I}_5(\la)\o270}+{\h{I}_4(\la)\o648}~.
}}
%\eqn\Wpertexp{
%W(\la,g_s)=2\h{I}_1(\la)+g_s^2{\h{I}_2(\la)\o3}+g_s^4\lf(-{\h{I}_4(\la)\o45}+
%{\h{I}_3(\la)\o36}\ri)+g_s^6\lf({2\h{I}_6(\la)\o945}-{\h{I}_5(\la)\o270}+{\h{I}_4(\la)\o648}\ri)
%+\cdots
%}
In general, $W_{\ell}(\la)$ is a linear combination of $\h{I}_k(\la)$
\eqn\WlinIk{
W_{\ell}(\la)=2\sum_{m=0}^{\ell-1} c_{m}
\h{I}_{\ell+m+1}(\la)~,
}
where $c_m$ is the coefficient of $g_s^{\ell-m}$ in $p_{\ell+m}(g_s)$.
Namely, $W_{\ell}(\la)$ is determined by certain coefficients
in $p_{\ell}(g_s),p_{\ell+1}(g_s),\cdots,p_{2\ell-1}(g_s)$.
This implies that the string loop correction at fixed order $\ell$
is calculable by a finite number of steps using \Pnint.

\subsec{Large $\la$ Limit}
From the expression \Wbuch, it is easy to find a
leading $g_s$-correction to the large $\la$ behavior of Wilson loop.
Recalling that the leading asymptotics of modified Bessel function $I_k(z)$
is independent of $k$
\eqn\Inasy{
I_k(z)\sim{1\o\rt{2\pi z}}e^z\quad(|z|\riya\infty)~,
}
\Wbuch\ becomes
\eqn\Wlargela{\eqalign{
W(\la,g_s)&\sim\rt{2\o\pi}\la^{-{3\o4}}e^{\rt{\la}}
\sum_{n=0}^\infty{g_s^n\o(\rt{\la})^n}p_n(g_s) \cr
&=\rt{2\o\pi}\la^{-{3\o4}}e^{\rt{\la}}
\exp\lf[g_sH\lf({g_s\o\rt{\la}}\ri)\ri]~.
}}
In the large $\la$ limit, the last factor is approximated by the
first term in the expansion of $g_sH(g_sz)$ in \sergsH.
Therefore, we find
\eqn\Wlacorre{
W(\la,g_s)\sim\rt{2\o\pi}\la^{-{3\o4}}
\exp\lf(\rt{\la}+{g_s^2\o6\rt{\la}}\ri)~.
}
This agrees with the result of \DrukkerRR.\foot{To compare with the expression in \DrukkerRR, note that
$g_s^2/6=(g_{\rm YM}^2)^2/96$.}
Note that the expression \Wlacorre\ is valid in the regime
\eqn\laregion{
\{\la\gg1\}\cap\{\la\gg g_s^2\}~.
}
In \DrukkerRR, it is argued that the $g_s$-correction found
in \Wlacorre\
is understood from the consideration of worldsheet moduli integral
in string theory on $AdS_5\times S^5$.

As argued in \DrukkerKX, when $g_s$ becomes of order $\rt{\la}$
the large $\la$ behavior of $W(\la,g_s)$ 
gets all order correction in $g_s^2/\la$
beyond the first order correction given in \Wlacorre.
In the bulk string theory side, this is computed by the 
DBI action of D3-brane \DrukkerKX
\eqn\WDBI{
W(\la,g_s)\sim\exp\lf[{\la\o2g_s}\sinh^{-1}\lf(g_s\o\rt{\la}\ri)
+\hf\rt{\la+g_s^2}\ri].
}
In the Yang-Mills/matrix model side, 
this is easily obtained by
the saddle point approximation of the contour integral \Wcont.
To reproduce this expression from the series expansion \Wbuch,
we have to keep sub-leading terms in the large $\la$
asymptotics of modified Bessel function.

\newsec{Expansion in Terms of Number of Holes}

The 't Hooft expansion \thooftexp\ is a double expansion
in $h$ and $\ell$, thus we can try to sum over $\ell$ first
and write $W$ in the form
\eqn\Winhexp{
W=\sum_{h=0}^\infty N^h{\cal F}_h(g_s)~,
}
where ${\cal F}_h(g_s)$ is the amplitude with fixed $h$
\eqn\Fhsum{
{\cal F}_h(g_s)=(4g_s)^h\sum_{\ell=0}^\infty{\cal W}_{h,\ell}g_s^{2\ell}~.
}
For later convenience, 
we included the factor $(4g_s)^h$ in $\la^h=(4g_sN)^h$
into the definition of ${\cal F}_h(g_s)$.
Usually we do not consider this form of expansion since it is not
so illuminating. 
However, one advantage of this expansion is that
we can study the analytic property of ${\cal F}_h(g_s)$ 
as a function of string coupling $g_s$, at least for fixed $h$.
This may give a clue to understand the $g_s$-dependence of circular
Wilson loop.

In order to write $W$ in the form \Winhexp, 
we use the second expression in \confW\
and write the confluent hypergeometric function as a summation
\eqn\Whypergeom{\eqalign{
W&=e^{-{g_s\o2}}\sum_{k=0}^\infty g_s^k{(N+1)(N+2)\cdots(N+k)
\o k!(k+1)!} \cr
&=e^{-{g_s\o2}}\sum_{k=0}^\infty{g_s^k\o (k+1)!}\prod_{j=1}^k\lf(1+{N\o j}\ri)
~.
}}
We would like to rewrite 
this as a power series in $N$.\foot{The $N$-dependent 
factor in the first line of \Whypergeom\ is known as the rising factorial.
This can be expanded in terms of 
the Stirling number of the first kind $s(n,m)$ 
$$(N+1)(N+2)\cdots(N+k)=\sum_{m=1}^ks(k+1,m+1)(-1)^{k-m}N^m$$
However, this expression is not so useful for our purpose,
so we will not use this.}
The zero-th order term is easily found by setting $N=0$ in 
\Whypergeom
\eqn\Fzero{
{\cal F}_0(g_s)=e^{-{g_s\o2}}\sum_{k=0}^\infty {g^k_s\o(k+1)!}
={2\o g_s}\sinh\lf({g_s\o2}\ri)~.
}
The higher $h$ term has the form
\eqn\Fhsum{
{\cal F}_h(g_s)=e^{-{g_s\o2}}\sum_{k=0}^\infty{g_s^k\o (k+1)!}
H^{(h)}_k~,
}
where the coefficient $H^{(h)}_k$ is determined recursively by
\eqn\hamH{\eqalign{
H^{(0)}_k=1,\qquad 
 H^{(h)}_k=\sum_{n=1}^k{1\o n}H_{n-1}^{(h-1)}~.
}}
In particular, the coefficient in the $h=1$ term is the
harmonic number: $H^{(1)}_k=\sum_{n=1}^k{1\o n}$.

Remarkably, it turns out that
${\cal F}_h(g_s)$ and ${\cal F}_{h+1}(g_s)$
are related by the following integral transformation
\eqn\Frecurs{
{\cal F}_{h+1}(g_s)=\int_0^1dt\,2\sinh\lf({g_s\o2}(1-t)\ri){\cal F}_h(g_st)~.
}
This is easily shown by expanding both sides in $g_s$ and using
the relation \hamH.
One can also write this relation \Frecurs\ 
in the form of convolution
\eqn\Fconv{
{\cal F}_{h+1}(g_s)={1\o g_s}({\cal B}*{\cal F}_h)(g_s)~,
}
where $*$ is defined by
\eqn\convdef{
(F*G)(x)=\int_0^xdy\, F(y)\,G(x-y)~.
}
The function ${\cal B}(g_s)$ appearing in \Fconv\ is given by
\eqn\Hkernel{
{\cal B}(g_s)=2\sinh\lf({g_s\o2}\ri)=g_s{\cal F}_0(g_s)~.
}
To summarize, one can increase the number of holes by one
by taking a convolution with ${\cal B}(g_s)$,
which is essentially given by the $h=0$ term.

As an illustration of the recursion relation,
let us consider the behavior of ${\cal F}_h(g_s)$ near $g_s=0$.
At the leading order in $g_s$, one can use the approximation
${\cal F}_0\sim 1$ and ${\cal B}\sim g_s$. Taking the convolution recursively,
one can easily find
\eqn\Ffirst{
{\cal F}_h(g_s)={g_s^h\o h!(h+1)!}+{\cal O}(g_s^{h+2})
}
Plugging this into \Winhexp, we recover the planar result 
$W_0(\la)=2\h{I}_1(\la)$.

Our relation \Fconv\ is not limited to the perturbative regime in $g_s$.
We are able to talk about the
analytic property of ${\cal F}_h(g_s)$. For $h=0$, it is clear
that ${\cal F}_0(g_s)$ given in \Fzero\ is analytic on the whole 
$g_s$-plane.
Let us look at the next term $h=1$. Setting $h=0$ in 
\Frecurs\ and evaluating the integral, we find
\eqn\Fone{
{\cal F}_1(g_s)={1\o g_s}\lf[e^{g_s\o2}{\rm Ein}(g_s)
+e^{-{g_s\o2}}{\rm Ein}(-g_s)\ri]
}
where ${\rm Ein}(x)$ denotes  the entire exponential integral\foot{The
entire exponential integral is related to the ordinary
exponential integral by
$${\rm E}_1(x)=\int_x^\infty{dt\o t}e^{-t}=-\ga-\log x+{\rm Ein}(x)~.$$
}
\eqn\Eindef{
{\rm Ein}(x)=\int_0^x{dt\o t}(1-e^{-t})=\sum_{k=1}^\infty
{(-1)^{k-1}x^k\o k!\,k}~.
}
Note that ${\cal F}_1(g_s)$ is regular at $g_s=0$.
The apparent
singularity at $g_s=0$ due to the overall factor
$1/g_s$ in \Fone\ is canceled by
the function inside the bracket which behaves as $g_s^2$ near
$g_s=0$. 
Therefore,
${\cal F}_1(g_s)$ vanishes linearly as $g_s\riya 0$, as expected from
the general form \Fhsum. 
Actually ${\cal F}_1(g_s)$ is
an entire function of $g_s$, since ${\rm Ein}(g_s)$ is entire 
as the name suggests.
By induction, one can argue that ${\cal F}_h(g_s)$ are analytic for all
$h$, since the integral \Frecurs\ is always convergent and there is no source
of singularity.

At least formally, our relation \Fconv\ suggests that
starting from $\la=0$
\eqn\lazero{
{\cal F}_0(g_s)=W(\la=0,g_s)~,
}
we can turn on $\la$ by successively applying the convolution
\eqn\Wcovhsum{
W(\la,g_s)
=\sum_{h=0}^\infty \lf({\la\o 4g_s}\ri)^h
\lf[{1\o g_s}\Big(g_s{\cal F}_0\Big)*\ri]^h{\cal F}_0(g_s)
}
In the next section, we will find a relation between
$g_s=0$ and $g_s\not=0$.

\newsec{Turning on $g_s$ from $g_s=0$}

\subsec{Formal Expression of $W$}
To see the relation between
$g_s=0$ and $g_s\not=0$,
let us start with the contour integral representation of
$W_0(\la)=W(\la,g_s=0)$
\eqn\Wzerocont{
W_0(\la)=2\oint_{z=0}{dz\o2\pi i}e^{\hf(\la z+{1\o z})}=2\h{I}_1(\la)~.
}
The key observation is that the differentiation of $W_0(\la)$ 
by $\la$
is equivalent to the insertion of $z$ in the contour
integral
\eqn\dellaz{
2{\del\o \del\la}\lrya z~.
}
By looking at the contour integral representation
of $W(\la,g_s)$ in \splitH, 
one immediately notices that
\eqn\Wgs{
W(\la,g_s)=\exp\lf[g_sH\lf(2g_s{\del\o \del\la}\ri)\ri]W_0(\la)~.
}
This relation can be also obtained from the expansion in terms
of $p_n(g_s)$ \Wbuch.
From \explaIsum, $\h{I}_k(\la)$ is given by
\eqn\Ikzint{
\h{I}_k(\la)=\oint_{z=0}{dz\o2\pi i}z^{k-1}e^{\hf(\la z+{1\o z})}~.
}
Again, using the correspondence \dellaz, one finds
that the derivative with respect to $\la$
increases the index of $\h{I}_k(\la)$
\eqn\Indel{
\lf(2{\del\o \del\la}\ri)^n\h{I}_k(\la)=
\h{I}_{k+n}(\la)~.
}
Using this, we can rewrite \Wbuch\ as
\eqn\Wdellasum{
W(\la,g_s)=2\sum_{n=0}^\infty\lf(2g_s{\del\o\del\la}\ri)^np_n(g_s)\h{I}_1(\la)
~,
}
which is equivalent to \Wgs\ from the definition of $p_n$ \pndef.

To summarize, we found a remarkable property of 
1/2 BPS circular
Wilson loop:
turning on the string coupling from zero coupling
amounts to acting
a certain differential operator
of $\la$ on the planar result $W_0(\la)$.

Although \Wgs\ is elegant, it is still a formal expression.
In particular, the differential operator appearing in \Wgs\
involves a derivative $\del_\la^n$ with arbitrarily large $n$.
Therefore, it might be better to interpret it as an
integral transform rather than a differential
operator. In the next subsection, we will consider
the Chern-Simons theory on $S^3$ as a simple example to see
if this interpretation works.

\subsec{Digression to the Chern-Simons Theory on $S^3$}
The partition function of $SU(N)$ Chern-Simons theory on $S^3$ 
is exactly known \WittenHF, and its 't Hooft expansion was
studied in \GopakumarKI. Via a geometric transition, this
is equivalent to the topological A-model on the resolved conifold.
The K\"{a}hler parameter of ${\Bbb P}^1$ is identified as the 
't Hooft coupling $t=g_sN$ of Chern-Simons theory.
The partition function looks like this:
\eqn\Zmacpert{
Z_{\rm conifold}(t,g_s)=Z_{\rm point}(g_s)Z_{\rm pert}(t,g_s)~,
}
where
\eqn\Mac{\eqalign{
Z_{\rm point}(g_s)&=\prod_{n=1}^\infty(1-e^{-ng_s})^{-n}~, \cr
Z_{\rm pert}(t,g_s)&=\prod_{n=1}^\infty(1-e^{-t-ng_s})^n~.
}}
%\eqn\topZ{
%Z_{\rm top}=Z_{\rm pert}Z_{\rm point}Z_{\rm np}
%}
%\eqn\Zpertpoint{\eqalign{
%&Z_{\rm pert}=\prod_{n=1}^\infty(1-e^{-t-ng_s})^n \cr
%&Z_{\rm point}=\prod_{n=1}^\infty(1-e^{-ng_s})^{-n} \cr
%&Z_{\rm np}=e^{-{t\o24}}\prod_{n=1}^\infty(1-e^{-ng_s})^{t\o g_s}
%}}
We are interested in the $t$-dependent part $Z_{\rm pert}(t,g_s)$.
The free energy has the following genus expansion
\eqn\loZpert{
\log Z_{\rm pert}(t,g_s)
={1\o g_s^2}F_0(t)-{t\o24}+F(t,g_s)~.
}
The first term is the genus zero term, the second term
is a part of the genus one term which is linear in $t$,
and the last term is a sum of $g\geq 2$ terms and a remaining part of $g=1$
term
\eqn\Fgenus{
F(t,g_s)=\sum_{g=1}^\infty g_s^{2g-2}{B_{2g}\o2g (2g-2)!}\Li_{3-2g}(e^{-t})~.
}
Here $\Li_k(z)$ denotes the polylogarithm
\eqn\polyL{
\Li_k(z)=\sum_{n=1}^\infty{z^n\o n^k}~.
}
%\eqn\Lzero{
%\Li_0(z)={z\o 1-z},\quad \Li_1(z)=-\log(1-z)
%}
From this definition, it follows that the function of $t$
appearing in the free energy \Fgenus\ satisfies
\eqn\derivLi{
\lf(-{\del\o \del t}\ri)^n\Li_k(e^{-t})=\Li_{k-n}(e^{-t})~.
}
One can immediately notice a similarity of this relation
and \Indel\ for $\h{I}_k(\la)$. From \derivLi, we can rewrite
\Fgenus\ as
\eqn\FdelK{
F(t,g_s)=\sum_{g=1}^\infty {B_{2g}\o2g (2g-2)!}
\lf(g_s{\del\o\del t}\ri)^{2g-2}\Li_1(e^{-t})
\equiv 12K\lf(g_s{\del\o\del t}\ri)F_1^{\rm inst}(t)~.
}
Here we introduced a function $K(x)$
\eqn\KxBg{
K(x)=\sum_{g=1}^\infty{B_{2g}\o2g (2g-2)!}x^{2g-2}~,
}
and a part of the genus one free energy written as a sum of
worldsheet instantons
\eqn\Foneinst{
F_1^{\rm inst}(t)={1\o12}\Li_1(e^{-t})=-{1\o12}\log(1-e^{-t})~.
}
\FdelK\ means that all loop free energy is obtained from
the $g=1$ term by acting a differential operator of 't Hooft coupling.

Again, the differential operator in \FdelK\ is of infinite order. 
One might try to define it by using the Fourier transform
of $K(x)$
\eqn\ideaK{
K(x)=\int dp \,\til{K}(p)e^{-ixp}~.
}
Then the action of $K(g_s\del_t)$ in \FdelK\ is given by 
a shift of $t$
\eqn\FtasKint{
F(t,g_s)=12 \int dp \,\til{K}(p)F_1^{\rm inst}(t-ig_sp)~.
}
Unfortunately, it turns out that
$K(x)$ is not Fourier-transformable.
However, there exists an expression like \ideaK\
with the integration region restricted to
the positive real axis $0\leq p\leq\infty$.
To see this, we use the integral representation
of Bernoulli number
\eqn\Bgint{
\int_0^\infty dp{2p^{2g-1}\o e^{2\pi p}-1}=(-1)^{g-1}
{B_{2g}\o2g}~.
}
Plugging this into \KxBg, $K(x)$ becomes
\eqn\Kphalf{\eqalign{
K(x)&=\sum_{g=1}^\infty\int_0^\infty dp{2p^{2g-1}\o e^{2\pi p}-1}
(-1)^{g-1}{1\o (2g-2)!}x^{2g-2} \cr
&=\int_0^\infty dp{2p\o e^{2\pi p}-1}\cos(xp)~.
}}
Finally, we arrive at an integral form of \FdelK
\eqn\Fgtpint{
F(t,g_s)=12 \int_0^\infty dp{p\o e^{2\pi p}-1}\Big[
F_1^{\rm inst}(t+ig_sp)+F_1^{\rm inst}(t-ig_sp)
\Big]~.
}

The failure of the existence of Fourier transform
of $K(x)$ is seen by writing \Kphalf\ as
\eqn\Ktotdel{
K(x)=\int_0^\infty dp\lf(p\coth(\pi p)\cos(xp)-{d\o dp}\sin(xp)\ri)~.
}
The second term cannot be dropped since it oscillates at the upper end
$p=\infty$. Let us ignore this term for the moment.
The first term in \Ktotdel\ can be extended to the whole 
$p$-axis
\eqn\Kfourier{
K(x)\sim \hf\int_{-\infty}^\infty dp\,p\coth(\pi p)e^{-ixp}~.
}
If we use this expression in \FdelK, we get
\eqn\Fcothint{
F(t,g_s)\sim 6\int_{-\infty}^\infty dp\,p\coth(\pi p)F_1^{\rm inst}(t-ig_sp)~.
}
By closing the contour of $p$-integral
and picking up the residues at $p=in~(n=1,2,\cdots)$,
we find
\eqn\sumFone{
F(t,g_s)\sim -12\sum_{n=1}^\infty nF_1^{\rm inst}(t+g_sn)
=\sum_{n=1}^\infty n\log(1-e^{-t-ng_s})
}
which is exactly the log of $Z_{\rm pert}(t,g_s)$ in the infinite
product form \Mac.

\subsec{Integral Transformation of $W_0(\la)$ to $W(\la,g_s)$}
Let us return to the circular Wilson loop case. In \Wgs, we found that
$g_s=0$ and $g_s\not=0$ are related by 
\eqn\Gxgs{
W(\la,g_s)=G(2g_s\del_\la,g_s)W_0(\la)
}
where $G(x,g_s)=e^{g_sH(x)}$.
As in the previous subsection, we will try to write it
in the form of  Fourier integral
\eqn\Ginp{
G(x,g_s)=\int dp\,\til{G}(p,g_s)e^{-ixp}~.
}
The Chern-Simons example suggests that the region of
$p$-integral does not necessarily
extend to the whole $p$-axis. Then \Gxgs\ is written as an integral transform
\eqn\WpintG{
W(\la,g_s)=\int dp\,\til{G}(p,g_s)W_0(\la-2ig_sp)~.
}
Note that the integral kernel $\til{G}(p,g_s)$ has an explicit
$g_s$-dependence whereas $\til{K}(p)$ in the Chern-Simons case does not.

Unfortunately, we do not know how to compute the kernel
$\til{G}(p,g_s)$ in a closed form. 
Therefore, in the following we will construct
the kernel perturbatively in $g_s$. By expanding $e^{g_sH(x)}$
as a power series in $g_s$
\eqn\expgHsum{
e^{g_sH(x)}=\sum_{n=0}^\infty {g_s^n\o n!}H(x)^n~.
}
we will rewrite $H(x)^n$ into the form \ideaK\ term by term.
From the explicit calculation of lower order terms,
we conjecture that $H(x)^n$ has the following
integral representation depending
on the parity of $n$: 
\eqn\Hkint{\eqalign{
&H(x)^{2k-1}=\int_0^\infty dp\,\sin(xp)G_{2k-1}(p)~, \cr
&H(x)^{2k}=\int_0^\infty dp\Big(1-\cos(xp)\Big)G_{2k}(p)~.
}}
Then \Gxgs\ is written in a form of integral transform 
\eqn\Weoint{
W(\la,g_s)-W_0(\la)=\int_0^\infty dp\Big[G_{-}(p,g_s)W_{-}(\la,g_sp)+
G_{+}(p,g_s)W_{+}(\la,g_sp)\Big]
}
where we defined
\eqn\Wpm{\eqalign{
W_{-}(\la,g_sp)&={W_0(\la+2ig_sp)-W_0(\la-2ig_sp)\o2i}~, \cr
W_{+}(\la,g_sp)&=W_0(\la)-{W_0(\la+2ig_sp)+W_0(\la-2ig_sp)\o2}~.
}}
and
\eqn\Apmexp{\eqalign{
G_{-}(p,g_s)=&\sum_{k=1}^\infty{g_s^{2k-1}\o(2k-1)!}G_{2k-1}(p) ~,\cr
G_{+}(p,g_s)=&\sum_{k=1}^\infty{g_s^{2k}\o(2k)!}G_{2k}(p) ~.
}}
In appendix C, we computed the integral kernel of $H(x)^n$
for first few orders. The result is
\eqn\Hint{\eqalign{
G_1(p)&=\Li_0(e^{-\pi p}) \cr
G_2(p)&=\hf p\,
\Li_0(e^{-\pi p})
+{\Li_1(e^{-\pi p})\o\pi}\cr
G_3(p)&={1\o 4}\lf[\Big(1-\hf p^2\Big)
\Li_0(e^{-\pi p})-3p{\Li_1(e^{-\pi p})\o\pi}-
6{\Li_2(e^{-\pi p})\o\pi^2}\ri] \cr
G_4(p)&={1\o 3!}\lf[
\Big(p-{p^3\o8}\Big)\Li_0(e^{-\pi p})+3\Big(1-\hf p^2\Big)
{\Li_1(e^{-\pi p})\o\pi}\ri.\cr
&\hskip41mm\lf.-\Big(3+{11p\o2}\Big){\Li_2(e^{-\pi p})\o\pi^2}
-16{\Li_3(e^{-\pi p})\o\pi^3}
\ri]~.
}}

Let us make a few comments on the properties of kernel
$G_{\pm}(p,g_s)$. As in the Chern-Simons case, the kernel has
poles on the imaginary $p$-axis.
Although we cannot close the contour
of $p$-integral, those poles are intimately
related to the analytic property of $W(\la,g_s)$
as a function of $g_s$.
One can observe that the lower order terms \Hint\ contains 
a piece
\eqn\foneexp{
\Li_0(e^{-\pi p})={1\o e^{\pi p}-1}
}
which has poles at
\eqn\polefone{
p_{\rm pole}=2in\qquad n\in{\Bbb Z}~.
}
At this pole, the shifted argument of $W_0(\la)$ in \Wpm\ is given by
\eqn\itatpole{
\la\mp 2ig_sp_{\rm pole} =\la\pm 4g_sn=4g_s(N\pm n)~.
}
Namely, the shift of $\la$ at $p_{\rm pole}$ is
equivalent to an integer shift of $N$.
This was also the case for the Chern-Simons theory.
We speculate that the existence of the poles at these
points reflects the underlying discreteness of $\la/4g_s$.

\newsec{Discussions}
In this paper, we studied the 't Hooft expansion of
1/2 BPS circular Wilson loop $W(\la,g_s)$
in ${\cal N}=4$ SYM and found its curious properties.

First, we found an operation which increases the number of
holes by one. We do not understand the physical meaning of
it. It would be nice to find a physical
origin of this relation.

Second, we found that $W(\la,g_s)$ is obtained by acting a
differential operator of $\la$ on $W(\la,0)$.
We also observed a similar relation
for the free energy $F(t,g_s)$ of Chern-Simons theory on $S^3$.
It might be the case that the similarity of 't Hooft expansion
of circular Wilson loop and Chern-Simons free energy
merely means that 1/2 BPS Wilson loop is essentially
a topological object.
Indeed, it is argued in \DrukkerRR\ that the whole
dependence of circular Wilson loop on $\la$ and $g_s$ 
is coming from the anomaly of conformal transformation
from straight line to circular loop.
We also suspect that this similarity is a consequence of
a general property of matrix model,
since both $W(\la,g_s)$ \refs{\EricksonAF,\DrukkerRR}
and $F(t,g_s)$ \refs{\MarinoFK,\AganagicWV} have matrix model representation.
In both cases, the amplitudes have the structure
\eqn\Alag{
A(t,g_s)=\sum_{n}g_s^nf_n(g_s)A_n(t),\quad \del_t^k A_n(t)=A_{n+k}(t)~.
} 
For the Chern-Simons case,
$f_n(g_s)={1+(-1)^n\o2}\chi_n$ with constant $\chi_n$.
It is interesting to see if this structure appears in other cases.

Besides similarity, there are some differences
between $W(\la,g_s)$ and $F(t,g_s)$.
The main source of difference
is the 't Hooft coupling dependence of worldsheet instanton action:
\eqn\Sinst{\eqalign{
\exp(-S_{\rm inst})&=\exp(\rt{\la})\hskip2.6mm:\,{\rm string~on~AdS}/{\cal N}=4~{\rm SYM}~,\cr
\exp(-S_{\rm inst})&=\exp(-t)\quad:{\rm topological~string}
/{\rm Chern{-}Simons~theory}~.
}}
For the AdS case, by $S_{\rm inst}$ we mean a regularized worldsheet action
for the minimal surface \DrukkerZQ.
Since the $g_s\not=0$ result is obtained by acting
$G(2g_s\del_\la,g_s)$ or $K(g_s\del_t)$ to the $g_s=0$ value, the 
$g_s$-dependence is closely related to
the 't Hooft coupling dependence of planar result.
In particular, the existence of $q$-expansion with $q=e^{-g_s}$
is related to the linear dependence of worldsheet instanton action
on 't Hooft coupling. Because of the form of the instanton action
$e^{\rt{\la}}$, the circular Wilson loop does not have 
a $q$-expansion as we saw in the text.

\appendix{A}{Contour Integral Representation of $W$}
In this appendix, we will show that the contour integral in \Wcont\
is equal to
the Laguerre polynomial expression of Wilson loop \Laggs.
First, we rescale the variable as $z\riya z/2g_s$. Then the integral \Wcont\
becomes
\eqn\Wrescale{\eqalign{
W&={1\o g_s}\oint_{z=0}{dz\o2\pi i}\exp\lf(Nz+{g_s\o2}\coth{z\o2}\ri)\cr
&={1\o g_s}e^{g_s\o2}\oint_{z=0}{dz\o2\pi i}\exp\lf(Nz+{g_s\o e^z-1}\ri)~.
}}
Here we used $\la=4g_sN$.
By the change of variable $e^z=1+w$, this is rewritten as
\eqn\Winwint{\eqalign{
W&={1\o g_s}e^{g_s\o2}\oint_{w=0}{dw\o2\pi i}(1+w)^{N-1}e^{g_s\o w}\cr
&={1\o g_s}e^{g_s\o2}\oint_{w=0}{dw\o2\pi i}\sum_{k=0}^{N-1}\lf(\matrix{N-1\cr
k}\ri)w^k\sum_{n=0}^\infty{g_s^n\o n!w^n} \cr
&=e^{g_s\o2}\sum_{k=0}^{N-1}\lf(\matrix{N-1\cr
k}\ri){g_s^k\o(k+1)!} \cr
&={1\o N}e^{g_s\o2}\sum_{k=0}^{N-1}\lf(\matrix{N\cr
k+1}\ri){g_s^k\o k!} ~.
}}
Recalling the definition of Laguerre polynomial
\eqn\Lagdef{
L_{N-1}^1(x)=\sum_{k=0}^{N-1}\lf(\matrix{N\cr
k+1}\ri){(-x)^k\o k!}~,
}
we can see that
the last expression of \Winwint\ is equal to the matrix model result 
\Laggs.

\appendix{B}{Some Properties of Buchholz Polynomials}
In this appendix, we summarize some useful properties of Buchholz polynomials
$p_n(a)$,
which are defined by
\eqn\Buch{
\exp\Big[aH(x)\Big]=\sum_{n=0}^\infty p_n(a)x^n~,
}
where $H(x)$ is 
\eqn\Hxsum{
H(x)=\hf\lf(\coth x-{1\o x}\ri)=\sum_{n=1}^\infty{x\o x^2+\pi^2 n^2}~.
}
Expanding the last expression around $x=0$,
we find that the Taylor coefficient $b_k$ of $H(x)$
\eqn\Hxcoefbk{
H(x)=\sum_{k=1}^\infty b_kx^{2k-1}
}
is given by the Riemann zeta function (or equivalently
by the Bernoulli number)
\eqn\bkdef{
b_k=(-1)^{k-1}{\zeta(2k)\o\pi^{2k}}={2^{2k-1}B_{2k}\o(2k)!}~.
}

Let us summarize some properties of $p_n(a)$ which follow
directly from the definition \Buch.
From the obvious relation $e^{(a+b)H}=e^{aH}e^{bH}$, it follows that
\eqn\delpn{
p_n(a+b)=\sum_{k=0}^{n}p_k(a)p_{n-k}(b)~.
}
Next, by taking the $x$-derivative on both sides of
\Buch\ and using \Hxcoefbk, we find a recursion relation
\eqn\prec{
p_n(a)={a\o n}\sum_{k=1}^{[{n+1\o2}]}(2k-1)b_k
p_{n+1-2k}(a)~.
}
This relation states that the $n^{\rm th}$ polynomial
$p_n(a)$ is determined by the lower order polynomials $p_{k}(a)~~(k<n)$.
Although this relation determines $p_n(a)$ recursively, it involves
many terms. A simpler relation between the consecutive neighbors
$p_n(a)$ and $p_{n-1}(a)$ can be found as follows.
From the definition \Hxsum, $H(x)$ satisfies the relation
\eqn\Hdel{
\hf{d\o d x}H(x)+{1\o x}H(x)=\qu-H(x)^2~.
}
This implies a differential equation for $e^{aH(x)}$
\eqn\gendel{
\lf({1\o2a}{\del\o\del x}+{1\o x}{\del\o\del a}\ri)e^{aH(x)}
=\lf(\qu-{\del^2\o\del a^2}\ri)e^{aH(x)}~,
}
which is equivalent to the following relation between $p_n(a)$ and
$p_{n-1}(a)$
\eqn\pnderiv{
{n\o2a}p_n(a)+p_n'(a)=\qu p_{n-1}(a)-p_{n-1}''(a)~.
}
One can easily integrate this equation and find the
integral form of this relation in the text
\Pnint.

The closed form expression of $p_n(a)$ is not known in the literature.
However, we can find some coefficients
of higher or lower powers of $a$.
The highest power term $a^n$ of $p_n(a)$ is determined by
the first term $b_1x=x/6$ in the expansion of $H(x)$ \Hxcoefbk.
Expanding the exponential $e^{ax/6}$, we find 
\eqn\pnfirst{
p_n(a)={a^n\o 6^nn!}+({\rm lower~order~terms})~.
}
Next, let us consider the lower order terms. From the expansion
of $e^{aH(x)}$ to the first order in $a$
\eqn\Hseries{
e^{aH(x)}=1+a\sum_{k=1}^\infty b_kx^{2k-1}+\cdots~,
}
it follows that the odd-order polynomial $p_{2k-1}(a)$
starts with the term $b_ka$. We can then fix the coefficient
of $a^2$ in the even-degree polynomial $p_{2k}(a)$
by using the recursion relation \pnderiv.
Setting $n=2k+1$ in \pnderiv\ and comparing the ${\cal O}(a^0)$
terms on both sides, we find that $p_{2k}(a)$ starts
with the term $-{2k+3\o4}b_{k+1}a^2$. We can repeat this process
and find some lower order terms as
\eqn\Pexp{\eqalign{
&p_{2k-1}(a)=b_ka+{1\o4\cdot 3!}\lf[{1\o2!}(2k+3)(2k+4)b_{k+1}+b_k\ri]a^3
+\cdots~,\cr
&p_{2k}(a)=-{2k+3\o4}b_{k+1}a^2 \cr
&\qquad-{1\o3!4!}\lf[{1\o 2^3}(2k+5)(2k+6)(2k+7)b_{k+2}+(2k+4)b_{k+1}\ri]a^4
+\cdots~.
}}

There is an alternative, more direct way to fix the coefficients
of $p_n(a)$. From the expansion
\eqn\easum{
\exp\Big[aH(x)\Big]=\sum_{k=0}^\infty {a^k\o k!}H(x)^k=\sum_{n=0}^\infty
x^np_n(a)~,
}
the coefficients of $a^k$ in $p_n(a)$ is given by
the Taylor coefficient
of $x^n$ in $H(x)^k/k!$. 
Therefore, all we need to know is the Taylor expansion 
of $H(x)^k$. For example, let us consider $H(x)^2$.
Using the last expression in \Hxsum, 
$H(x)^2$ is written as
\eqn\Hxsq{\eqalign{
H(x)^2&=x^2\sum_{n=1}^\infty\sum_{m=1}^\infty{1\o
(x^2+\pi^2n^2)(x^2+\pi^2m^2)} \cr
&=x^2\sum_{n=1}^\infty{1\o(x^2+\pi^2n^2)^2}+x^2\sum_{n\not=m}
{1\o(x^2+\pi^2n^2)(x^2+\pi^2m^2)} \cr
&=x^2\sum_{n=1}^\infty{1\o(x^2+\pi^2n^2)^2}
+x^2\sum_{n\not=m}\lf({1\o x^2+\pi^2n^2}-{1\o x^2+\pi^2m^2}\ri)
{1\o\pi^2(m^2-n^2)} \cr
&=x^2\sum_{n=1}^\infty{1\o(x^2+\pi^2n^2)^2}+x^2\sum_{n=1}^\infty
{1\o x^2+\pi^2n^2}{3\o2\pi^2n^2}~.
}}
In the last step, we used the relation
\eqn\summn{
\sum_{m\geq 1,m\not=n}{1\o m^2-n^2}={3\o4n^2}~.
}
Finally, expanding the last expression in \Hxsq, we reproduce
the coefficient of $a^2$ in $p_{2k}(a)$ obtained in \Pexp\
by the recursion relation.
As one can easily see, this approach becomes 
extremely cumbersome as the power of $H(x)^k$ increases.

\appendix{C}{Integral Representation of $H(x)^k$}
In this appendix, we will rewrite $H(x)^k$ 
as a Fourier-like integral.

Let us start with the $k=1$ case. Using the expansion of $H(x)$
\Hxcoefbk\ with
coefficient given in \bkdef, we find
\eqn\Hone{\eqalign{
H(x)&=\sum_{k=1}^\infty(-1)^{k-1}\sum_{n=1}^\infty
{1\o(\pi n)^{2k}}x^{2k-1} \cr
&=\int_0^\infty dp\sum_{k=1}^\infty(-1)^{k-1}
{t^{2k-1}\o (2k-1)!}x^{2k-1} \sum_{n=1}^\infty e^{-\pi n p}\cr
&=\int_0^\infty dp\,\sin(xp)\,\Li_0(e^{-\pi p})~.
}}

Next consider $k=2$. The Taylor coefficient of $x^{2k}$ in $H(x)^2/2!$ is 
the coefficient of $a^2$ in $p_{2k}(a)$ \Pexp.  
Therefore, we find
\eqn\Htwo{\eqalign{
H(x)^2&=-\hf\sum_{k=1}^\infty(2k+3)b_{k+1}x^{2k} \cr
&=\hf\sum_{k=1}^\infty(-1)^{k-1}(2k+3)\sum_{n=1}^\infty
{1\o (\pi n)^{2k+2}}x^{2k} \cr
&=\int_0^\infty dp
\sum_{k=1}^\infty(-1)^{k-1}\lf[{2k+1\o2}{p^{2k+1}\o(2k+1)!}\sum_{n=1}^\infty
e^{-\pi n p}+
{p^{2k}\o (2k)!}\sum_{n=1}^\infty{1\o\pi n}e^{-\pi n p}\ri]x^{2k}\cr
&=\int_0^\infty dp\,\Big(1-\cos(xp)\Big)\lf[\hf p\,\Li_0(e^{-\pi p})
+{1\o\pi}\Li_1(e^{-\pi p})\ri]~.
}}
In the third step, we split $2k+3$ into $(2k+1)+2$ and
introduced the $p$-integral with different power, $p^{2k+1}$ or $p^{2k}$. 

The computation of the higher power of $H(x)$ is similar (but tedious).
We can find the Taylor coefficient of $H(x)^n/n!$
by looking at the $a^n$ term in \Pexp. 
Using the coefficients in \Pexp,
we obtain the kernel of $H(x)^3$ and $H(x)^4$ written
in the text \Hint.

\listrefs
\bye